\documentclass[10pt,final]{IEEEtran}

\usepackage[final]{graphicx}
\usepackage{amsmath,mathrsfs,amssymb}
\newcommand{\bc}{\begin{center}}
\newcommand{\ec}{\end{center}}
\newcommand{\bdm}{\begin{displaymath}}
\newcommand{\edm}{\end{displaymath}}
\newcommand{\beq}{\begin{equation}}
\newcommand{\eeq}{\end{equation}}
\newcommand{\bfl}{\begin{flushleft}}
\newcommand{\efl}{\end{flushleft}}
\newcommand{\bt}{\begin{tabbing}}
\newcommand{\et}{\end{tabbing}}

\usepackage{here}
\numberwithin{equation}{section} \allowdisplaybreaks
\begin{document}


\title{Finding Sequence Features in Tissue-specific Sequences}

\author{ Arvind~Rao, Alfred~O.~Hero~III, David~J.~States, James~Douglas~Engel
\thanks{Arvind~Rao and Alfred~O.~Hero,~III are with the Departments of Electrical
Engineering and Computer Science, and Bioinformatics at the
University of Michigan, Ann Arbor, MI-48109, email: [ ukarvind,
hero] @umich.edu}
\thanks{James~Douglas~Engel is with the Department of Cell and Developmental Biology at the University of Michigan, Ann Arbor, MI-48109.}
\thanks{David~J.~States is with the Departments of Bioinformatics and Human Genetics at the University of Michigan, Ann Arbor, MI-48109.}
}

\maketitle

\begin{abstract}
The discovery of motifs underlying gene expression is a challenging
one. Some of these motifs are known transcription factors, but
sequence inspection often provides valuable clues, even discovery of
novel motifs with uncharacterized function in gene expression.
Coupled with the complexity underlying tissue-specific gene
expression, there are several motifs that are putatively responsible
for expression in a certain cell type. This has important
implications in understanding fundamental biological processes, such
as development and disease progression. In this work, we present an
approach to the principled selection of motifs (not necessarily
transcription factor sites) and examine its application to several
questions in current bioinformatics research.

There are two main contributions of this work: Firstly, we introduce
a new metric for variable selection during classification , and
secondly, we investigate a problem of finding  specific sequence
motifs that underlie tissue specific gene expression. In conjunction
with the SVM classifier we find these motifs and discover several
novel motifs which have not yet been attributed with any particular
functional role (eg: TFBS binding motifs). We hypothesize that the
discovery of these motifs would enable the large-scale investigation
for the tissue specific regulatory potential of any conserved
sequence element identified from genome-wide studies.

Finally, we propose the utility of this developed framework to not
only aid discovery of discriminatory motifs, but also to examine the
role of any motif of choice in co-regulation or co-expression of
gene groups.
\end {abstract}


\IEEEpeerreviewmaketitle

\section{Introduction}

\begin{figure}
\centerline{\includegraphics[width=3.5in,height=1.5in]{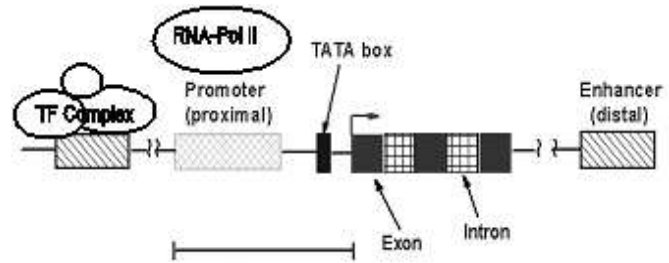}}
\caption{Schematic of Transcriptional
Regulation.}\label{fig:transcription}
\end{figure}

Understanding the mechanisms underlying regulation of tissue
specific gene expression is still a challenging question that does
not have a satisfactory explanation. While all mature cells in the
body have a complete copy of the human genome, each cell type only
expresses those genes it needs to carry out its assigned task. This
includes genes required for basic cellular maintenance (often called
"house keeping genes") and those genes whose function is specific to
the particular tissue type the cell belongs to. Gene expression by
way of transcription is the process of generation of messenger RNA
(mRNA) from the DNA template representing the gene. It is the
intermediate step before the generation of functional protein from
messenger RNA. During gene expression (Fig.
\ref{fig:transcription}), transcription factor proteins are
recruited at the proximal promoter of the gene as well as at
sequence elements (enhancers/silencers) which can lie several
hundreds of kilobases from the gene's transcriptional start site.

The combination of the basal transcriptional machinery at the
promoter coupled with the transcription factor complexes at  the
distal regulatory elements are collectively involved in directing
tissue specific expression of genes. Some of the common features of
these distal regulatory elements are:

\begin{itemize}
\item
Non-coding elements: Distal regulatory elements are non-coding and
can eitehr be intronic or intergenic region on the genome. Hence
previous models of gene finding are not applicable herein. Moreover,
with over 98\% of the annotated genome being non-coding the precise
localization of regulatory elements that underlie tissue-specific
gene expression is a challenging problem.
\item
Distance/orientation independent: An enhancer can act from great
genomic distances (hundreds of kilobases) to regulate gene
expression in conjunction with the proximal promoter, possibly via a
looping mechanism. They can lie upstream or downstream of the actual
gene.
\item
Promoter dependent: Since the action at a distance of these elements
involves the recruitment of transcription factors that direct
tissue-specific gene expression, the promoter that they interact
with is critical.
\end{itemize}

Though there are instances where a gene harbors tissue specific
activity at their promoter itself, there is considerable interest to
examine the long range elements (LREs) for a detailed understanding
of their regulatory role in gene expression during biological
processes like organ development and disease progression
\cite{Kleinjan2005}. An in-silico examination of the sequence
properties of such LREs would result in computational strategies to
find novel LREs genomewide that govern tissue specific expression
for any gene of interest.

Our primary question in this regard is: are there any discriminating
sequence properties of such LRE elements that determine tissue
specific gene expression - more particularly, are there any sequence
motifs in such regulatory elements that can aid discovery of new
elements with similar potential for tissue-specific regulation
genomewide \cite{RPscore}. We remind the reader that a similar
approach was used in gene finding algorithms, wherein sequence
features of exons are examined to facilitate the discovery of novel
genes. In popular gene finding algorithms \cite{Burge1997},
discriminatory motifs are identified from annotated genes and
non-coding regions. Such an approach can be extended to this
situation too wherein we look for motifs discriminating regulatory
and neutral non-coding elements.
In this work, we explore the possible existence of such
discriminating sequence motifs in two kinds of biologically relevant
regulatory sequences:

\begin{itemize}
\item
\underline{Promoters of tissue specific genes}: Before the
widespread discovery of long-range regulatory elements (LREs) it was
hypothesized that promoters governed gene expression entirely. There
is substantial evidence for the binding of tissue-specific
transcription factors at the promoters of expressed genes. This
suggests that, in spite of newer information implicating the role of
LREs, promoters might also have interesting motifs that govern
tissue-specific expression, that are potentially relevant to the
discovery of new LREs de-novo.

Another practical reason for the examination of promoters is that
their locations (and genomic sequences) are more unambiguously
delineated on genome databases (like UCSC or Ensembl). Sufficient
data (http://symatlas.gnf.org) on the expression of genes is
publicly available for analysis.

We set up the motif discovery as a feature extraction problem from
these tissue-specific promoter sequences and then build  a support
vector machine (SVM) classifier to classify new promoters into
specific and non-specific categories based on the identified
sequence features (motifs).
Using the SVM classifier algorithm we are able to accurately
classify 70\% of tissue specific genes based upon their upstream
promoter region sequences alone.

\item
\underline{Known long range regulatory elements (LRE) motifs}: To
analyze the motifs in LRE elements, there is a need for an
experimental dataset of elements that confer tissue-specific
expression in some eukaryotic animal model. For our purpose, we
examine the results of this approach on other new data source of
interest - the Enhancer Browser at \emph{http://enhancer.lbl.gov}
which has results of expression of ultraconserved genome elements in
transgenic mice \cite{EnhancerBrowser}. An examination of these
ultraconserved enhancers is useful for the extraction of
discriminatory motifs to distinguish these regulatory elements from
non-regulatory (neutral) ones. Here the results indicate that upto
90\% of the sequences can be correctly classified using these
identified motifs.
\end{itemize}

We note that some of the identified motifs might not be
transcription factor binding motifs, and would need to be
functionally characterized. This is an advantage of our method -
instead of constraining ourselves by the degeneracy present in TF
databases (like TRANSFAC/JASPAR), we look for all sequences of a
fixed length.


\section{Contributions}

From microarray expression data,
(\cite{Shimizu2006},\cite{Stoeckert2005}) proposes an approach to
assign genes into tissue specific and non-specific types using an
entropy criterion. They use the variation in expression and its
divergence from ubiquitous expression (uniform distribution across
all tissue types) to make this assignment. From this assignment,
several features like CpG island density, frequency of transcription
factor motif occurrence, can be examined to discriminate these two
subgroups. However, a denovo examination of 'every' sequence feature
in sequence and its subsequent interpretation has not been pursued
adequately. Other work has explored the existence of key motifs
(transcription factor binding sites) in the promoters of
tissue-specific genes \cite{Genomatix}.

For the purpose of identifying discriminative motifs from the
training data (tissue-specific promoters or LREs), our approach is
two-pronged:
\begin{itemize}
\item
\emph{Variable selection}: We first find those sequence motifs that
can discriminate between tissue-specific and non-specific elements.
In machine learning, this is a feature selection problem. Here,
these features are the counts of sequence motifs in these training
sequences. Without loss of generality, we can use six-nucleotide
motifs (hexamers) as the motif features. This is based on the
observation that most transcription factor binding motifs have a 5-6
nucleotide core sequence with degeneracy at the ends of the motif.
Another reason is that we need to find a length that captures
biologically meaningful information as well as does not lead to an
unduly large search space. Even though our motivation for choosing
hexamers was independent, a similar setup has been referred to in
(\cite{DrosophilaFeatureDiff}, \cite{PromFind}). We find that $4^6$
possibilities (from hexamer sequences) yields good performance
without being unduly costly, computationally. The overall presented
approach, however, does not depend of this choice of motif length
and can be scaled depending on biological intuition.

In the first part of the work in this paper, we present a novel
feature selection approach (based on a new information theoretic
quantity called directed information - DTI) that can be applied to
such learning problems.
\item
\emph{Classifier design}: After discovering key discriminating
motifs from the above DTI step, we proceed to build a SVM classifier
that separates the samples between the two classes (specific and
non-specific) from this feature space. One distinction we make is
that the class label is a perceptual abstraction (based on our
intuition and experience). The feature space (hexamer counts) is a
physical/measurement space - this is what we can look for.
\end{itemize}

Apart from the novel feature selection approach, our contributions
to bioinformatics methodology are outlined below:

From the identified motifs, we ask several related questions:.
\begin{itemize}
\item
Are there any common motifs identified from tissue-specific
promoters and corresponding enhancers, both underlying expression in
the same tissue?. To answer this, we examine brain-specific
promoters and enhancers.
\item
Do these motifs correlate with known motifs (transcription factor
binding sites),
\item how useful are these motifs in predicting new regulatory elements?.
\end{itemize}

 This work differs from
that in (\cite{DrosophilaFeatureDiff}, \cite{PromFind}), in several
aspects. We present the novel DTI based feature selection procedure
as part of an overall unified framework to answer several questions
in bioinformatics, not limited to finding discriminating motifs
between two classes of sequences. Particularly, one of the
advantages of the proposed approach is the examination of any
particular motif as a potential discriminator between two classes.
Also, this work accounts for the notion of tissue-specificity of
promoters/enhancers (in line with more recent work in
\cite{CNSGata3},\cite{EnhancerBrowser},\cite{Shimizu2006},\cite{Stoeckert2005},\cite{Khandekar2004}).
This is clarified further in the Results (Sections: $XI$ and $XII$).
After solving the main problem posed in this work viz,
identification of tissue specific motifs from annotated sequences
(promoter/LREs), we will then examine some other related problems
which would benefit from such an analysis.

\section{Rationale}

Some of the approaches to finding motifs relevant to certain classes
with respect to examining common motifs driving gene regulation is
summarized in (\cite{Kreiman2004}, \cite{Fraenkel2006}). The most
common approach is to look for TFBS motifs (TRANSFAC / JASPAR) that
are statistically over-represented based on a binomial or Poisson
model in the promoters of the co-expressed genes. This assumes a
parametric form on the background density of distribution of motifs
in promoters.

In our situation, we set-up the problem of discriminative motif
discovery as a word-document classification problem. Having
constructed two groups of genes for analysis, tissue specific ('ts')
and non-tissue specific ('nts') - we are interested to find hexamer
motifs which are most discriminatory between these two classes. Our
goal would be to make this set of motifs as small as possible - i.e.
to achieve maximal class partitioning with the smallest feature
subset. This is the classic feature-selection problem.

Several metrics have been proposed that can find features which have
maximal association with the class label. In information theory,
mutual information is a popular choice. This is a symmetric
association metric and does not resolve the direction of dependency
(i.e if features depend on the class label or vice versa). It is
important to find features that are induced by the class label. When
we capture features relevant to a data sample, feature selection
implies selection (control) of a feature subset that maximally
captures the underlying character (class label) of the data . We
have no control over the label (a purely perceptual
characterization).

With this motivation, we propose the use of a new
metric, termed "Directed Information" (DTI) for finding such a
discriminative hexamer subset. Subsequent to feature selection,
we design a Support Vector Machine (SVM) classifier to classify
 sequences that are tissue-specific or not.

As expected, the input to such an approach would be a gene
promoter - motif frequency table (Table~\ref{motif_promoter_mtx}).
The genes relevant to each class are identified from tissue
microarray analysis, and the frequency table is built by parsing the
gene promoters for the presence of each of the $4^6 = 4096$ possible
hexamers.

\begin{table}[!h!b!t]
\centering
\begin{tabular}{llllll}
Ensembl Gene ID & AAAAAA    & AAAAAG  & AAAAAT  & AAAACA \\
ENSG00000155366 & 0      & 0   & 1   & 4 \\
ENSG000001780892 & 6   & 5 &  5 & 6 \\
ENSG00000189171 & 1  &  2 &  1 &  0 \\
ENSG00000168664 & 6  &  3 &  8 & 0 \\
ENSG00000160917 & 4  &  1 &  4 &  2 \\
ENSG00000163655 & 2  &  4 &  0 & 1 \\
ENSG000001228844 &  8   & 6 &  10 & 7 \\
ENSG00000176749 & 0  &  0 &  0 & 0 \\
ENSG00000006451 & 5  &  2 &  2  & 1
\end{tabular}
\caption{The 'motif frequency matrix' for a set of gene-promoters.
The first column is their ENSEMBL gene identifiers and the other 4
columns are the motifs. A cell entry denotes the number of times a
given motif occurs in the upstream (-2000 to +1000bp from TSS)
region of each corresponding gene.}\label{motif_promoter_mtx}
\end{table}

\section{Methods}
Below we present our approach to find promoter specific or
enhancer-specific motifs.

\subsection{Promoter motifs:}
\subsubsection{Microarray Analysis} Raw microarray data was obtained
from the Novartis Foundation (GNF)
[\emph{http://symatlas.gnf.org/}]. Data was normalized using RMA
from the Bioconductor packages for R [\emph{cran.r-project.org/}].
Following normalization, replicate samples were averaged together.
Only 25 tissue types were used in our analysis including: Adrenal,
Amygdala, Brain, Caudate Nucleus, Cerebellum, Corpus Callosum,
Cortex, Dorsal Root Ganglion, Heart, HUVEC, Kidney, Liver, Lung,
Pancreas, Pituitary, Placenta, Salivary, Spinal Cord, Spleen,
Testis, Thalamus, Thymus, Thyroid, Trachea, and Uterus.

In order to classify genes as tissue specific or not, we had to
define what tissue-specific expression means in our context. The
notion of tissue-specificity is fairly ambiguous. We
define a gene as being tissue specific if it is expressed in no more
than three tissue types. We also defined non-tissue specific genes
as those being expressed in at least 22 of the 25 tissue types we
examined. A binary assignment method was employed to determine if a
gene was highly expressed in a given tissue type. In this method,
any gene whose expression level was at least two-fold greater than
the median expression level for the tissue type was considered to be
highly expressed and was assigned a score of one. Genes not meeting
this requirement were given an assignment of zero for that
particular tissue type. Using this approach a single numerical
summary could be achieved for every gene (across all tissue types).
This value could be used to find how many genes were highly
expressed in most tissue types and those that were only expressed in
a few. We note that, 
this also allows ample flexibility to a biologist to examine the
genes that she is interested in.


Suppose there are $N$ genes, $g_1,g_2,\ldots,g_N$ and $T$ tissue
types (in GNF: $T = 25$), we construct a $N\times T$  tissue
specificity matrix : $M = [0]_{N\times T}$. For each gene $g_i, 1
\leq i \leq N$, let $g_{i,[0.5T]} = median(g_{i,k}), \forall k \in
{1,2,\ldots,T}$. Define, each entry
$M_{i,k}$ as, \\

\[ M_{i,k} = \left\{ \begin{array}{ll}
1 & \mbox{if $g_{i,k} \geq 2g_{i,[0.5T]}$};\\
0 & \mbox{otherwise}.\end{array} \right. \]\\

Now consider the $N$ dimensional vector $m_i = \sum _{k=1}^T
M_{i,k}, 1 \leq i \leq N$ i.e. summing all the columns of each row.
Plot the inter-quartile range of $'m'$. For the gene indices $'i'$
lying in quartile 1 (=3), label as 'ts', and for indices in quartile
4 (= 22), label as 'nts'.

With this approach, a total of 1924 probes representing 1817 genes
were classified as tissue specific, while 2006 probes representing
2273 genes were classified as non-tissue specific. For our studies,
we consider genes which are either heart-specific or brain-specific.
From the tissue-specific genes obtained in the above approach, we
find 55 gene promoters that are brain specific and 118 gene
promoters that are heart-specific. The objective in this work is to
find motifs that are responsible for brain/heart specific expression
and possibly correlate them (atleast a subset) with binding profiles
of known transcription factor binding motifs.

\subsubsection{Sequence Analysis} Genes associated with candidate
probes were identified using the Ensembl Ensmart
[\emph{http://www.ensembl.org/}]. For each gene, we extracted 2000bp
upstream and 1000bp down-stream upto the start of the first exon
relative to their reported transcriptional start site in the Ensembl
Genome Database (Release 37). The relative counts of each of the
$4^6$ hexamers were computed within each gene-promoter sequence of
the two categories ('ts' and 'nts') - using the \emph{'seqinr'}
library in the R environment to parse these sequences and obtain the
frequency of occurrence (counts) of each hexamer in a sequence.. A
paired t-test was performed between the relative counts of each
hexamer between the two expression categories ('ts' and 'nts') and
the top 1000 significant hexamers having a p-value less than
$10^{-6}$ were selected ($\overrightarrow{\textbf{X}} =
X_1,X_2,\ldots,X_{1000}$). The relative counts of these hexamers was
computed again for each gene individually. This resulted in two
hexamer-gene co-occurrence matrices, each with $N_{train}$ rows of
genes and $M = 1000$ columns of hexamers - one for the 'ts' class
and the other for the 'nts' class. We note that $N_{train} =
min(S_{ts},S_{nts})$ with $S_{ts}$ being the number of positive
training ('ts') samples and $S_{nts}$ being the number of negative
training ('nts') samples. This is done to avoid bias problems during
learning.

\subsection{LRE motifs:}
To analyze long range elements which confer tissue specific
expression, we examine the Mouse Enhancer database
(\textit{http://enhancer.lbl.gov/}). Briefly, this database has a
list of experimentally validated ultraconserved elements which have
been tested for tissue specific expression in transgenic mice
\cite{EnhancerBrowser}. This database can be searched for a list of
all elements which have expression in a tissue of interest. In our
case, we consider expression in tissues relating to the developing
brain. We note that according to the experimental protocol, the
various regions are cloned upstream of a heat shock protein
promoter, not adhering to the idea of promoter specificity in
tissue-specific expression. Though this is of concern in that there
is loss of some gene-specific information, we work with this data
since we are more interested in tissue expression and also because
there is paucity of public enhancer-dependent data .

This database also has a collection of ultraconserved elements that
do not have any transgenic expression in-vivo. This is the
neutral/background set of data which corresponds to the 'nts'
(non-tissue specific class) during feature selection  and classifier
design.

As in the above (promoter) case, we can parse these sequences (sixty
two enhancers for brain-specific expression) for the absolute counts
of the $4096$ hexamers, build a co-occurrence matrix ($N_{train} =
62$) and then use t-test p-values to find the top 1000 hexamers
($\overrightarrow{\textbf{X'}}=X'_1,X'_2,\ldots,X'_{1000}$) that are
maximally different between the two classes (brain-specific and
brain non-specific).

The next three sections clarify the preprocessing, feature selection
and classifier design steps to mine these co-occurrence matrices for
hexamer motifs that are strongly associated with the class label. We
note that though we illustrate this work using two class labels, the
method can be extended in a straightforward way to the multi-class
problem.

\section{Preprocessing}

From the above, we now have $N_{train} \times 1000$ co-occurrence
matrices each for the tissue-specific and non-specific data, both
for the promoter and enhancer sequences. Before proceeding to the feature
(hexamer motif) selection step, we would need to normalize the
counts of the $M = 1000$ hexamers in each training sample. For this,
we can obtain an interquartile range of the hexamer counts in each
gene, and create equivalent co-occurrence matrices of dimension
$N_{train} \times 1000$ where each entry is the quantile membership
of the hexamer count. In this work, we use a ($K = 4$)-quantile
label
assignment. 

In this co-occurrence matrix, let $gc_{i,k}$ represents the absolute
count of the $k^{th}$ hexamer, $k \in {1,2,\ldots,M}$ in the
$i^{th}$ gene. Then, for each gene $g_{i}$, the quantile labeled
matrix has $g_{i,k} = l$ if $gc_{i,[\frac{l-1}{K}M]} \le gc_{i,k} <
gc_{i,[\frac{l}{K}M]}$

We can now construct matrices of dimension $N_{train} \times 1001$ for
each of the specific and non-specific training samples. Each matrix
would contain the quantile label assignments for the $1000$ hexamers
$(X_i, i \in (1,2,\ldots,1000))$, as stated above, and the last
column would have the class label ($Y = -1/+1$). These two matrices
are then integrated into one composite \emph{training data matrix}
of dimension $2N_{train} \times 1001$.

\section{Directed Information and Feature Selection }

\begin{figure}[h]
\centerline{\includegraphics[width=3in,height=2.2in]{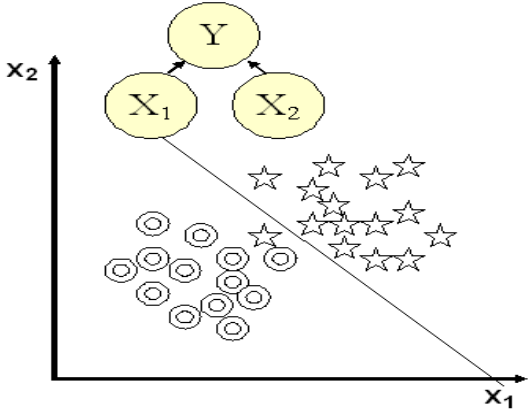}}
\caption{ Causal Feature discovery, adapted from \cite{Guyon2004}.
Here the variables $X_1$ and $X_2$ discriminate $Y$.
}\label{fig:dti_var}
\end{figure}

The primary goal in feature selection is to find the minimal subset
of features (from hexamers: $X_{i,1:1000}$) that lead to maximal
discrimination of the class label ($Y_i \in (-1/+1)$), using each of
the $i \in (1,2,\ldots,2N_{train})$ genes for training. We are
looking for a subset of the variables ($X_{i,1},\ldots,X_{i,1000}$)
which are directionally associated with the class label ($Y_{i}$).
These hexamers putatively influence/induce the class label (Fig. \ref{fig:dti_var}). 
As can be seen from [\emph{http://research.ihost.com/cws2006/}],
there is considerable interest in using causality to solve this
problem by discovering dependencies from the given data. Our
interpretation \cite{Guyon2004}, is to find features (in
\emph{measurement} space) that induce the class
label (in \emph{perceptual} space). 

%

There has been a lot of previous work exploring the feasibility of
using mutual information (MI) as a method to infer such conditional
dependence/influence between features and class labels
\cite{MI_feature} by exploring the structure of the joint
distribution of motif frequency profiles from the count matrix.
However, this metric is undirected and does not resolve whether the
hexamers induce the class label or vice-versa. This resolution is
essential since one can only control the physical/measurement
feature space, whereas the perceptual space (class label) remains
the same. Hence, the only freedom we have during learning is: which
features, among the ones that we collect, are maximally
representative of the class label or data type. The absence of such
a 'directed' information theoretic metric has prevented us from
exploitation of the full potential of information theory. We thus
examine the Directed Information (DTI) criterion as a potential
metric to the explicit inference of feature influence. This enables
us to uncover any meaningful relationship between features ($X_i$)
and class label ($Y$). In a regression (state-space) framework, the
measurements are
the state variables and the class label is the observation. 



\begin{figure}
\centerline{\includegraphics[width=3.8in,height=3in]{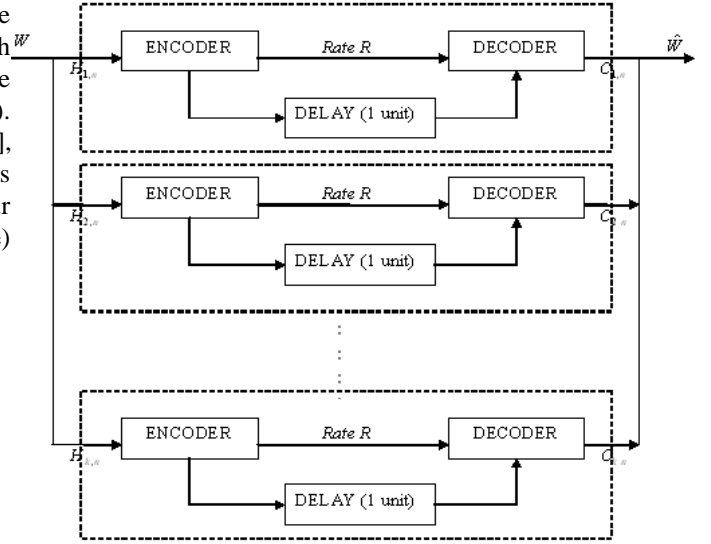}}
\caption{Directed Information setup under Source coding with
feedforward }\label{fig:DTI_source}
\end{figure}

\begin{figure}
\centerline{\includegraphics[width=3.8in,height=3in]{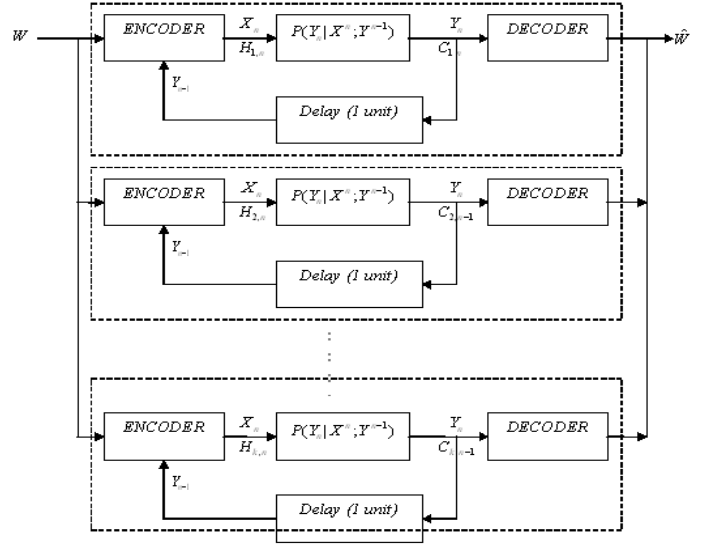}}
\caption{Directed Information setup under Channel Coding with
feedback }\label{fig:DTI_channel}
\end{figure}

A brief background about DTI is in order: Directed Information comes
from a rich literature regarding capacity of channels with feedback
or understanding rate-distortion theory in source coding with
feedforward \cite{Massey1990}. Source coding and channel coding are
information-theoretic duals, and DTI is useful to characterize
source or channel behavior when the information being transferred is
correlated \cite{Pradhan2007}.

%

The relationship between MI and DTI is given by,
\\ MI: \quad $I(X^N ; Y^N) = \sum_{i=1}^N
I(X^n;Y_i|Y^{i-1}) = I(X^N \rightarrow Y^N)+I(0Y^{N-1} \rightarrow
X^N).$
\\ DTI: \quad $I(X^N \rightarrow Y^N) = \sum_{i=1}^N I(X^i;Y_i|Y^{i-1})$.

Just like the case where we maximize $I(X;Y)$ in channel coding, or
minimize $I(X;Y)$ for source coding in cases without feedback, we
maximize/minimize $I(X \rightarrow Y)$ in the corresponding cases
with feedback/feedforward. This feature selection problem for the
$i^{th}$ training instance becomes one of identifying which hexamer
($k \in (1,2,\ldots,4096)$) has the highest $I(X_{i,k} \rightarrow
Y_i)$ or the lowest $I(Y_i \rightarrow X_{i,k})$ if looked from the
source or channel coding perspectives respectively.

In our case, we treat the hexamers ($X_i$) like messages and the
class labels ($Y$) as the reconstruction after transmission and are
interested in asking which messages transmit maximum information to
the reconstructed versions. These are the features which maximally
influence the class label. 
At each instant (in time or training iteration), the learning
algorithm uses the previous associations between the hexamer
features ($X_{k,1},X_{k,2},\ldots, X_{k,i-1}$) and corresponding
class labels ($Y_1,Y_2,\ldots,Y_{i-1}$) to predict a class label
($\hat{Y}_i$) for the current hexamer feature instance ($X_i$).
Following the concept of feedforward, the true class label $Y_i$ is
then given to the algorithm (after the prediction $\hat{Y}_i$) is
made. Based on the prediction ($\hat{Y}_i$) and the true class label
$Y_i$, the algorithm will learn the hexamer-label associations up
until the instance it has just observed (i.e. the $i^{th}$ instant),
and so on. It does this for each hexamer $'k'$ in the feature
vector, and finds the hexamer which has maximum association within
this sequential learning paradigm (Figs \ref{fig:DTI_source} and
\ref{fig:DTI_channel}). Each dotted box represents the DTI based
feature learning for each hexamer. Thus, this is thought of $M=1000$
operations in parallel. The training iterations proceed as long as
it takes to achieve a certain accuracy of classification. We would
expect that DTI needs lesser number of features than MI to achieve
the same classification rate.


The DTI (for a lag of 1) - which is a measure of the causal
dependence between two random processes $X_i =
[X_{1,i},X_{2,i},\ldots X_{n,i}]$, (with $X_{j,i}$ = quantile label
for the frequency of hexamer $i \in (1,2,\ldots,1000)$ in the
$j^{th}$ training sequence) and $Y = [Y_1,Y_2,\ldots,Y_n]$ being the
corresponding class labels ($-1/+1$), is given by [6]:
\begin{align*}
I(X_i^N \rightarrow Y^N) = \sum_{n=1}^{N}I(X_i^n;Y_n|Y^{n-1})
\tag{$1$}
\end{align*}
Here, $N$ is $2N_{train}$, the total number of training data class
labels. As already known, the mutual information $I(X; Y) =
H(X)-H(X|Y)$, with $H(X)$ and $H(X|Y)$ being the Shannon entropy of
$X$ and the conditional entropy of $X$ given $Y$, respectively.
Using this definition of mutual information, the Directed
Information simplifies to,

\begin{gather*}
I(X^N \rightarrow Y^N) = \sum_{n=1}^N [H(X^n|Y^{n-1})-H(X^n|Y^{n})] \\
= \sum_{n=1}^N \{ [H(X^n,Y^{n-1})-H(Y^{n-1})]-
[H(X^n,Y^{n})-H(Y^{n})]\} \tag{$2$}
\end{gather*}

Using (2), the Directed information is expressed in terms of
individual and joint entropies of $X$ and $Y$. For the purpose of
entropy estimation, we can adopt several approaches. In this work we
examine two different ways to estimate the directed information from
the hexamer-sequence frequency matrix.

%
%
%
%

The first way is to bin each frequency vector into $L$ quantiles.
Thus, within the $i^{th}$ sequence (promoter/LRE), we can find the
distribution of the hexamers within the sequence and bin them into
the appropriate quantile. The value/entry in each cell of Table I is
$l \in (1,2,\ldots,L)$. The last element in the data vector is the
class label ($-1/+1$). Hence, it is now straightforward to find the
marginal and joint distributions for the $k^{th}$ hexamer
($X_{i,k}$) and the class label ($Y_i$).

An alternate method to find the joint information of the random
variables $X^N$ and $Y^N$ uses the Darbellay-Vajda algorithm
\cite{Darbellay-Vajda1999}. From $(2)$, we also have,

\begin{eqnarray*}
I(X^N \rightarrow Y^N) = \sum_{n=1}^N [H(X^n|Y^{n-1})-H(X^n|Y^{n})] \\
 = \sum_{n=1}^N [H(X^n)-I(X^n;Y^{n-1})]-[H(X^n)-I(X^n;Y^n)]\\\label{eq:03}(\text{using }
I(X^n;Y^n)=H(X^n)-H(X^n|Y^n))\\  = \sum_{n=1}^N [I(X^n;Y^n) -
I(X^n;Y^{n-1})] \\ = \sum_{n=1}^N [I(X^n;Y^n) -
I(X^n;0Y^{n-1})]\label{eq:04}
\end{eqnarray*}

In the above expressions, 
the mutual information between two random variables in the sum
($I(X^n;Y^n)$ and $I(X^n;0Y^{n-1})$) can be estimated using a
non-parametric adaptive binning procedure (\cite{SignalProcMag2006},
\cite{Darbellay-Vajda1999}) by iterative partitioning of the
observation space until conditional independence is achieved within
and between partitions. This method lends itself to a tree based
partitioning scheme and can be used for entropy estimation even for
a moderate number of samples in the observation space of the
underlying probability distribution. Several such algorithms for
adaptive density estimation have been proposed
(\cite{WilettNowak2004},\cite{NemenmanBialek2002},\cite{Paninski2003},
\cite{ErikLearnedMiller2003}) and can find potential application in
this procedure. Because of the higher performance guarantees in
using this procedure as well as the relative ease of implementation,
we use the Darbellay-Vajda approach for entropy (and information)
estimation in our methodology.

%

Both these methods (equiquantization and Darbellay-Vajda) provide a
way to estimate the true DTI between a given hexamer and the class
label for the entire training set. Feature selection comprises of
finding all those hexamers ($X_i$) for which $I(X_i \rightarrow Y)$
is the highest. From the definition of DTI, we know that $0 \le
I(X_i \rightarrow Y) \le I(X_i;Y) < \infty$. To make a meaningful
comparison of the strengths of association between different
hexamers and the class label, we need to find a normalized score
(Sec:$VII$) to rank the DTI values. Another point of consideration
is that we need to ask how significant the DTI value is compared to
a null distribution on the DTI value (i.e. what is the chance of
finding the DTI value by chance from the series $X_i$ and $Y$). This
is done using confidence intervals after permutation testing (Sec:
$IX$).


%
%
%
%
%
%
%
%
%
%

\section*{A normalized DTI measure}
In this section, we derive an expression for a 'normalized DTI
coefficient'. This is useful for a meaningful comparison across
different criteria during network inference. For now, we will
compare the network influences as inferred from normalized DTI and
CoD \cite{CoD2003}. In this section, we use $X$,$Y$, $Z$ for $X^N$,
$Y^N$ and $Z^N$ interchangeably, i.e $X \equiv X^N$, $Y \equiv Y^N$,
and $Z \equiv Z^N$.

By the definition of DTI, we can see that $0 \leq I(X^N \rightarrow
Y^N) < I(X^N;Y^N) < \infty$. The normalized measure $\rho_{DTI}$
should be able
to map this large range ($[0,\infty]$) to $[0,1]$. 

We recall that the multivariate canonical correlation is given by
\cite{Gubner2006}:
$\rho_{X^N;Y^N}=\Sigma_{X^N}^{-1/2} \Sigma_{X^N;Y^N} \Sigma_{Y^N}^{-1/2}$\\
and this is normalized having eigenvalues between 0 and 1. We also
recall that, under a Gaussian distribution on $X^N$ and $Y^N$, the
joint entropy $H(X^N;Y^N) = -\frac{1}{2}\ln (2\pi e)^{2N}
|\Sigma_{X^N Y^N}|$, where $|A|$ is the determinant of matrix $A$,
$\Sigma$ denotes the covariance matrix.


Thus, for $I(X^N;Y^N) = H(X^N) + H(Y^N) - H(X^N;Y^N)$, the
expression for mutual information, under jointly Gaussian
assumptions on $X^N$ and $Y^N$, becomes,\\
$I(X;Y) = -\frac{1}{2} \ln (\frac{|\Sigma_{X^N
Y^N}|^2}{|\Sigma_{X^N}|. |\Sigma_{Y^N}|}) =
-\frac{1}{2}\ln(1-\rho_{X^N;Y^N}^2)$. Hence, a straightforward
transformation is normalized MI, $\rho_{MI}= \sqrt{1-e^{-2 I(X;Y)}}
= \sqrt{1-e^{-2\sum_{i=1}^N I(X^N;Y_i|Y^{i-1})}}$ . A connection
with \cite{John1989b}, can thus be immediately seen.

Both of these will be normalized between $(0,1)$ and will give a
better absolute definition of dependency that does not depend on the
unconditioned MI. We will use this definition of normalized
information coefficients for the present set of simulation studies.

For constructing a normalized version of the DTI, we can extend this
approach , from (\cite{Geweke1982}, \cite{John1989b}). Consider
three random vectors \textbf{X}, \textbf{Y} and \textbf{Z}, each of
which are identically distributed as
$\mathcal{N}(\mu_X,\Sigma_{XX})$, $\mathcal{N}(\mu_Y,\Sigma_{YY})$,
and $\mathcal{N}(\mu_Z,\Sigma_{ZZ})$ respectively. We also have,\\

$(\textbf{X},\textbf{Y},\textbf{Z}) \sim \mathcal{N} \left[\left(
\begin{array}{c}
\mu_X  \\
\mu_{Y}  \\
\mu_{Z}  \end{array} \right) , \left(
\begin{array}{ccc}
\Sigma_{XX} & \Sigma_{XY} & \Sigma_{XZ} \\
\Sigma_{YX} & \Sigma_{YY} & \Sigma_{YZ} \\
\Sigma_{ZX} & \Sigma_{ZY} & \Sigma_{ZZ} \end{array} \right)\right]$ \\

Their partial correlation $\delta_{YX|Z}$ is given by,\\
$\delta_{YX|Z} = \sqrt{\frac{a_2^2}{a_1 a_3}}$
with, \\
$a_1 = \Sigma_{YY}-\Sigma_{YZ}\Sigma_{ZZ}^{-1}\Sigma_{ZY}$,
$a_2 =\Sigma_{YX}-\Sigma_{YZ}\Sigma_{ZZ}^{-1}\Sigma_{ZX}$, \\
$a_3 = \Sigma_{XX}-\Sigma_{XZ}\Sigma_{ZZ}^{-1}\Sigma_{ZX}$\\

Recalling results from conditional Gaussian distributions, these can
be denoted by: $a_1 = \Sigma_{Y|Z},a_2 = \Sigma_{XY|Z}$ and $a_3 =
\Sigma_{X|Z}$.

Thus, $\delta_{YX|Z} =
\Sigma_{Y|Z}^{-1/2}\Sigma_{XY|Z}\Sigma_{X|Z}^{-1/2}$. Extending the
above result from the mutual information to the directed information
case, we have, 
$\rho_{DTI}= \sqrt{1-e^{-2\sum_{i=1}^N I(X^i;Y_i|Y^{i-1})}}$.

To once again clarify, we recall the primary difference between MI
and DTI, (note the superscript on X)\\ MI: \quad $I(X^N ; Y^N) =
\sum_{i=1}^N I(X^N;Y_i|Y^{i-1}) = I(X^N \rightarrow Y^N) +
I(0Y^{N-1} \rightarrow X^N)$.
\\ DTI: \quad $I(X^N \rightarrow Y^N) = \sum_{i=1}^N I(X^i;Y_i|Y^{i-1})$.

\section*{Kernel Density Estimation (KDE)}

The goal in density estimation is to find a probability density
function $\hat{f}(z)$ that approximates the underlying density
$f(z)$ of the random variable $Z$. Under certain regularity
conditions, the kernel density estimator $\hat{f}_h(Z)$ at the point
$x$ is given by $\hat{f}_h(Z) =
\frac{1}{nh}\sum_{i=1}^{n}K(\frac{z_i-z}{h})$, with $n$ being the
number of samples $z_1,z_2,\ldots,z_n$ from which the density is to
be estimated, $h$ is the bandwidth of a kernel $K(\bullet)$ that is
used during density estimation.

A Kernel density estimator at $z$ works by weighting the samples (in
$(z_1,z_2,\ldots,z_n)$) around $z$ by a kernel function (window) and
counts the relative frequency of the weighted samples within the
window width. As is clear from such a framework, the choice of
kernel function $K(\bullet)$ and the bandwidth $h$ determines the
fit of the density estimate.

Some figures of merit to evaluate various kernels are the asymptotic
mean integrated squared error (AMISE), bias-variance characteristics
and region of support \cite{HastieTibshirani2002}. It is preferred
that a kernel have a finite range of support, low AMISE and a
favorable bias-variance tradeoff. The bias is reduced if the kernel
bandwidth (region of support) is small, but has higher variance
because of a small sample size. For a larger bandwidth, this is
reversed (ie large bias and smaller variance). Under these
requirements, the Epanechnikov kernel has the most of these
desirable characteristics - i.e. a compact region of support, the
lowest AMISE compared to other kernels, and a favorable bias
variance tradeoff \cite{HastieTibshirani2002}.

The Epanechnikov kernel is given by:
\begin{displaymath}
K(u)=\frac{3}{4} (1-u^2) I(\left\vert u \right\vert \le 1).
\end{displaymath}
with $I(\bullet)$ being the indicator function conveying a window of
width spanning $[-1,1]$ centered at $0$. An optimal choice of the
bandwidth is $h = 1.06 \times \hat{\sigma}_z \times n^{-1/5}; $,
following [\cite{Silverman1997}]. Here $\hat{\sigma}_z$ is the
standard error from the bootstrap DTI samples
$(z_1,z_2,\ldots,z_n)$.

Hence the kernel density estimate for the bootstrapped DTI (with $n
= 1000$ samples), $Z
\triangleq \hat{I}_B(X^N \rightarrow Y^N)$ becomes,\\
$\hat{f}_h(Z) = \frac{1}{nh}\sum_{i=1}^{n}\frac{3}{4}
[1-(\frac{z_i-z}{h})^2] I(\left\vert \frac{z_i-z}{h} \right\vert \le
1)$ with $h \approx 2.67\hat{\sigma}_z$ and $n=1000$. We note that
$\hat{I}_B(X^N \rightarrow Y^N)$ is obtained by finding the DTI for
each random permutation of $X$, $Y$ time series, and performing this
permutation $B$ times.

\section{Bootstrapped Confidence Intervals}
Since we do not know the true distribution of the DTI estimate, we
find an approximate confidence interval for the DTI estimate
($\hat{I}(X^N \rightarrow Y^N)$), using bootstrap above
\cite{EffronTibshirani1994}.  

We denote the cumulative distribution function (over the Bootstrap
samples) of $\hat{I}(X^N \rightarrow Y^N)$ by $F_{\hat{I_B}(X^N
\rightarrow Y^N)}(\hat{I_B}(X^N \rightarrow Y^N))$. 
 Let the mean of the bootstrapped null distribution be
$I_B^{*}(X^N \rightarrow Y^N)$. We denote by $t_{1-\alpha}$, the
$(1-\alpha)^{th}$ quantile of this distribution i.e.
\{$t_{1-\alpha}: P([\frac{\hat{I_B}(X^N \rightarrow Y^N)-I_B^{*}(X^N
\rightarrow Y^N)}{\hat{\sigma}}] \leq t_{1-\alpha}) = 1-\alpha$\}.
Since we need the real $\hat{I}(X^N \rightarrow Y^N)$ to be
significant and as close to 1, we need $\hat{I}(X^N \rightarrow Y^N)
\geq [I_B^{*}(X^N \rightarrow Y^N)+ t_{1-\alpha} \times
\hat{\sigma}]$, with $\hat{\sigma}$ being the standard error of the
bootstrapped distribution, \\
$\hat{\sigma} = \sqrt{\frac{[\Sigma_{b=1}^B \hat{I_b}(X^N
\rightarrow Y^N)-I_B^{*}(X^N \rightarrow Y^N)]^2}{B-1}}$; $B$ is the
number of Bootstrap samples.


%
%
%
\section{Support Vector Machines}
From the top $"d"$ features identified from the ranked list of
features having high DTI with the class label, a hyperplane linear
classifier in these $"d"$ dimensions is designed. A Support Vector
Machine (SVM) is a hyperplane classifier which operates by finding a
maximum margin linear hyperplane to separate two different classes
of data in high dimensional ($D >"d"$) space. Our training data has
$N_{train}$ pairs $(x_1,y_1),(x_2,y_2),\ldots,
(x_{N_{train}},y_{N_{train}})$, with $x_i \in \mathscr{R}^D$ and
$y_i \in \{-1,+1\}$.

An SVM is a maximum margin hyperplane classifier in a non-linearly
extended high dimensional space. For extending the dimensions from
$d$ to $D > d$, a radial basis kernel is used. 

The objective is to minimize $||\beta||$ in the
hyperplane $\{x: f(x)=x^T \beta +\beta_0\}$, subject to \\
$y_i(x^{T}_{i}\beta+\beta_0) \geq 1-\xi_i \forall i$, $\xi_i \geq 0,
\sum \xi_i \leq$ constant \cite{HastieTibshirani2002}.

%

\section{Summary of Overall Approach}

Our proposed approach is as follows. Here, the term 'sequence' can
pertain to either tissue specific promoters or LRE sequences.
\begin{itemize}
\item
Parse the sequence to obtain the relative counts/frequencies of
occurrence of the hexamer in that sequence and build the
hexamer-sequence frequency matrix. The \emph{'seqinr'} package in R
is used for this purpose. This is done for all the sequences in the
specific (class $+1$) and non-specific (class $-1$) categories. The
matrix thus has $N_{train}$ rows and $4^6 = 4096$ columns.
\item
Preprocess the obtained hexamer-sequence frequency matrix by finding
the quantile labels for each hexamer within the $i^{th}$ sequence.
We now have a hexamer-sequence matrix where the cell $(i,j)$ has the
quantile label of the $j^{th}$ hexamer in the $i^{th}$ sequence.
This is done for all the $N (= S_{ts}+S_{nts})$ training sequences
consisting of examples from the $-1$ and $+1$ class labels.
\item
Build two submatrices corresponding to the two class labels. Thus
one matrix will contain the hexamer-sequence quantile labels for the
positive training examples and the other matrix is for the negative
training examples.
\item
To pick the hexamers that are most different between the positive
and negative training examples, we perform a paired t-test for each
hexamer. Rank all the corresponding t-test p-values from lowest to
highest and take the top $1000$ hexamers. These correspond to the
$1000$ hexamers that are most different distributionally (in mean)
between the positive and negative training samples. 
We note that the t-test requires the same number of samples in the
positive ($S_{ts}$ samples) and negative ($S_{nts}$ samples)
training set. Hence, we consider $N_{train} = min(S_{ts},S_{nts})$
examples for each of the positive and negative training cases.
Another way to resolve this problem is to find the symmetrized KL
divergence/ Jensen-Shannon divergence between the hexamer
distributions of the positive and negative examples. This step is
only necessary to reduce the computational complexity of the overall
procedure - finding the DTI between each of the 4096 hexamers and
the class label is very expensive.
\item
For the top $K = 1000$ hexamers which are most significantly
different between the positive and negative training examples, we
proceed to find $I(X_{i,k} \rightarrow Y_{i})$ and $I(X_{i,k}
\rightarrow Y_i)$ for each of the $k \in (1,2,\ldots,K)$ hexamers.
The entropy terms in the directed information and mutual information
expressions can be found either from the equidistant binning
approach or the Darbellay-Vajda approach. Since the goal is to
maximize  $I(X_{i,k} \rightarrow Y_{i})$ or minimize $I(Y_i
\rightarrow X_{i,k})$, we can rank them in descending and ascending
order, respectively. Using the procedure of Section.VII, the raw DTI
values can be converted into their normalized versions.
\item
We also find the significance of the DTI estimate obtained in the
step above. Thus if we set a threshold of $0.05$ significance, we
can take every hexamer whose DTI is $0.05$ significant with respect
to its bootstrapped null distribution (using kernel density
estimation), and rank the hexamers by decreasing DTI value. The top
$"d"$ hexamers in this ranked list can be used for classifier (SVM)
training.
\item
We now train the Support Vector Machine classifier (SVM) on the top
$"d"$ features from the ranked DTI list(s). For comparison with the
MI based technique, we use the hexamers which have the top $"d"$
(normalized) MI values. We can now plot the accuracy of the trained
classifier as a function of the number of features ($d$). As we
gradually consider higher $"d"$, we move down the ranked list.
\end{itemize}


\section{Results}


%
%
%
%
%

\subsection{Tissue specific promoters}
We use DTI to find discriminating hexamers that underlie brain
specific and heart specific expression.

\begin{figure}[h]
\centerline{\includegraphics[width=3.3in,height=2.8in]{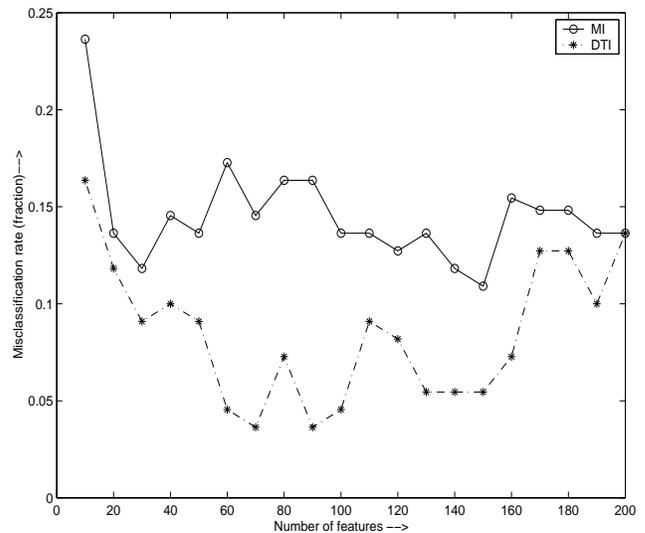}}
\caption{Misclassification accuracy for the MI vs. DTI case (brain
promoter set)}\label{fig:brain_ts}
\end{figure}

\begin{figure}[h]
\centerline{\includegraphics[width=3.3in,height=2.8in]{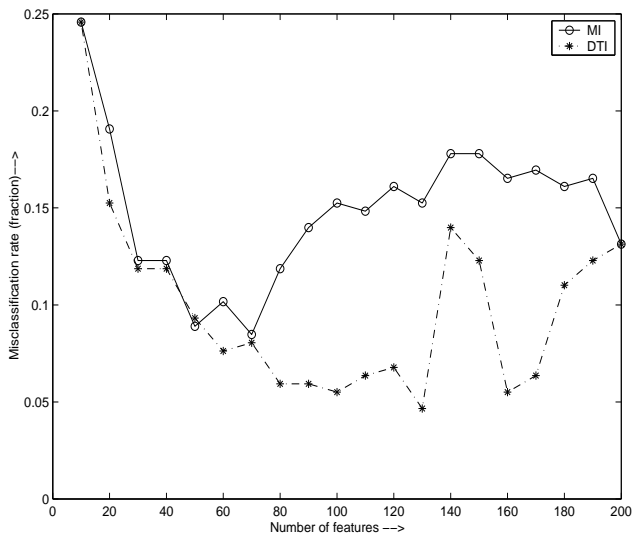}}
\caption{Misclassification accuracy for the MI vs. DTI case (heart
promoter set)}\label{fig:heart_ts}
\end{figure}

Results for the MI and DTI methods are given below
(Figs.\ref{fig:brain_ts} and \ref{fig:heart_ts}). The plots indicate
the cross-validated misclassification accuracy (ideally $0$) for the
data as the number of features using the metric (DTI or MI) is
gradually increased. We can see that for any given classification
accuracy, the number of features using DTI is less than the
corresponding number of features using MI. This translates into a
lower misclassification rate for DTI-based feature selection. We
observe that as the number of features $"d"$ is increased the
performance of MI is the same as DTI.

Some of the key motifs in the heart and brain case are given in
Table II. Wherever possible, we indicate if the motif corresponds to
a known transcription factor binding motif. We note that just
because a motif corresponds to a transcription factor binding site
(TFBS), it does not imply that the TF is functional in brain or
heart. It might however, be useful to do
focused experiments to check their functional role. 
\begin{table}[h]
\centering
\begin{tabular}{lcccc}
\hline
    Brain         & Heart         & Brain         \\
    -promoters    & promoters     & enhancers  \\
   \hline
   Ahr-ARNT & Pax2 & HNF-4                         \\
   Tcf11-MafG & Tcf11-MafG & Nkx     \\
   c-ETS &       XBP1 & AML1                 \\
   FREAC-4 &       Sox-17 & c-ETS                  \\
   T3R-alpha1 & FREAC-4 & Elk1                              \\
   \end{tabular}
\caption{Comparison of high ranking motifs (by DTI) across different
data sets.}
\end{table}

\subsection{Enhancer DB}
\begin{figure}[h]
\centerline{\includegraphics[width=3.3in,height=2.8in]{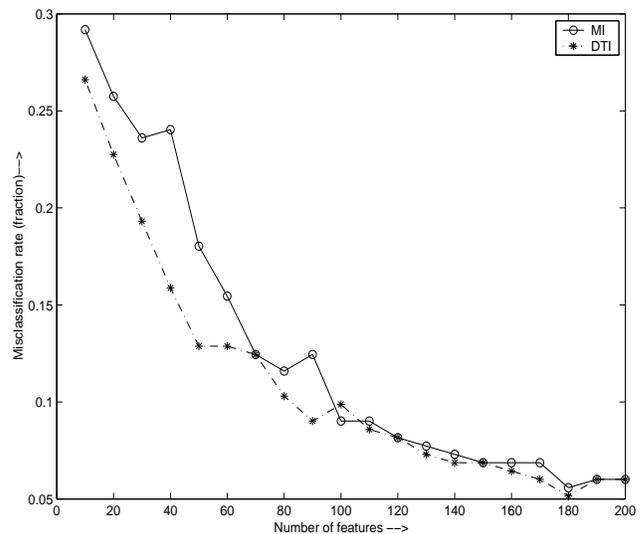}}
\caption{Misclassification accuracy for the MI vs. DTI case (brain
enhancer set)}\label{fig:brain_enh}
\end{figure}

We examine all the brain -specific regulatory elements profiled in
the EnhancerDB database (\emph{http://http://enhancer.lbl.gov/}) for
discriminating motifs. Again, the plot of misclassification accuracy
vs. number of features in the MI and DTI scenarios are indicated in
Figs. \ref{fig:brain_ts}-\ref{fig:brain_enh}.


%
%
%

Some of the top ranking motifs from this dataset are also shown in
Table II.

\section{Other Applications}
We now proceed to show other related applications wherein the DTI
based learning framework is useful. Compared to other approaches we
can investigate  the role of \emph{'any'} motif (possibly a
transcription factor binding motif) both in sequence or via
expression data. We illustrate this via an example.

\subsection{}
Suppose we are interested in the transcription factors that regulate
\textit{Gata3} gene expression. This gene has expression in the
developing kidney, central nervous systems and hematopoietic cell
differentiation. In concordance with the established framework of
transcription presented in Section I, recruitment of TFs happens at
both the proximal promoter as well as long-range regulatory
elements. A common approach to find functional TFBSes from the
promoter sequence is to look for phylogenetically conserved TF
binding sites in the promoter sequence among species in which
\textit{Gata3} is involved in the same biological process (here
kidney development).

\begin{figure}[h]
\centerline{\includegraphics[width=3.4in,height=1.5in]{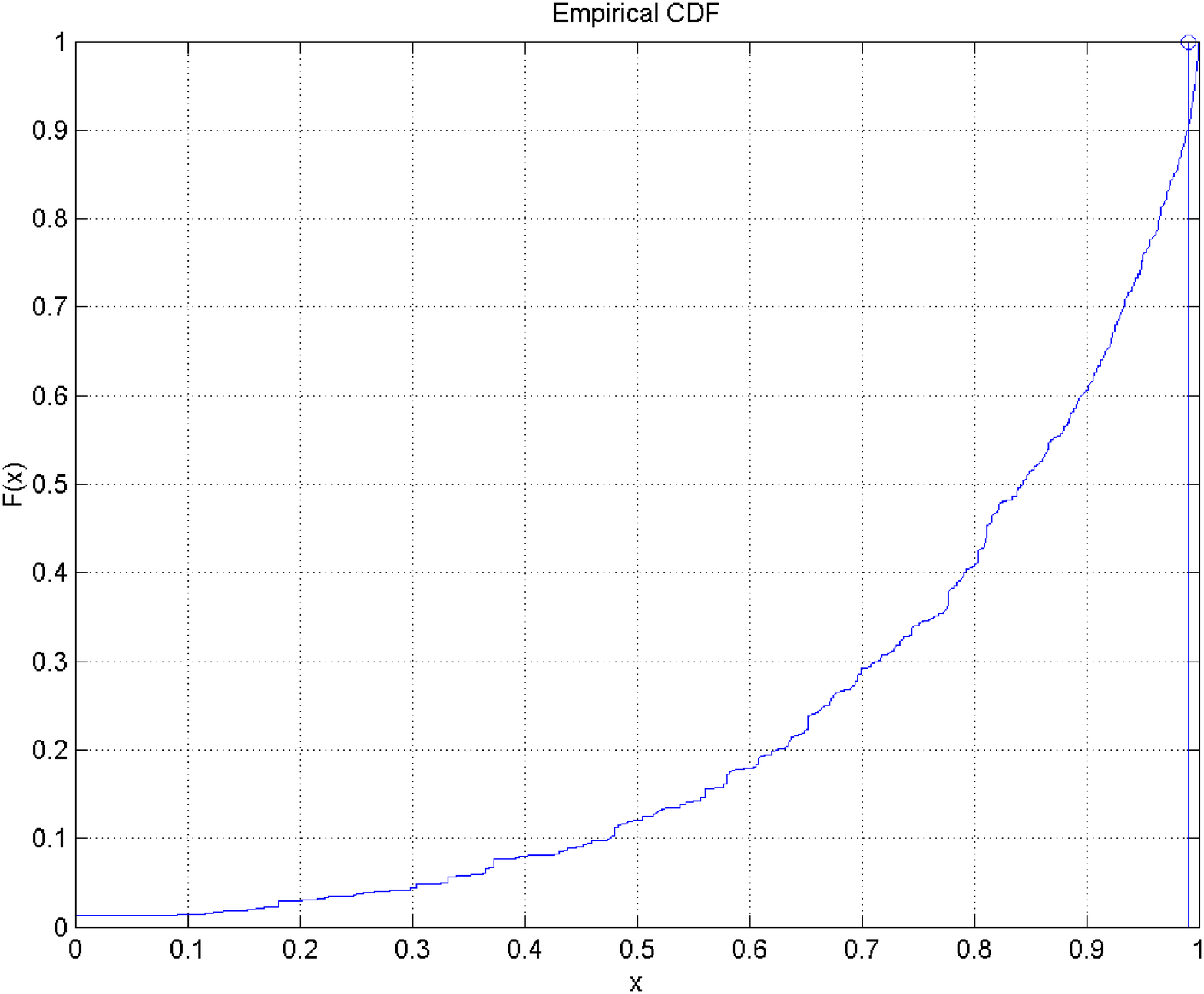}}
\caption{Cumulative Distribution Function for bootstrapped
$I(\textit{Pax2} \rightarrow \textit{Gata3})$. True $\hat{I}
(\textit{Pax2} \rightarrow \textit{Gata3}) =
0.9911$.}\label{fig:cdf}
\end{figure}

As shown above, this DTI is seen to be significant and strong. DTI
also enables the integration of microarray time series data
expression (from the developing kidney) of the \textit{Pax2} gene to
ask if there is any influence from \textit{Pax2} to \textit{Gata3}.
This is not discussed here but some preliminary work is
available in \cite{icassp2006}. 

\begin{figure}
\centerline{\includegraphics[width=3.4in,height=1.5in]{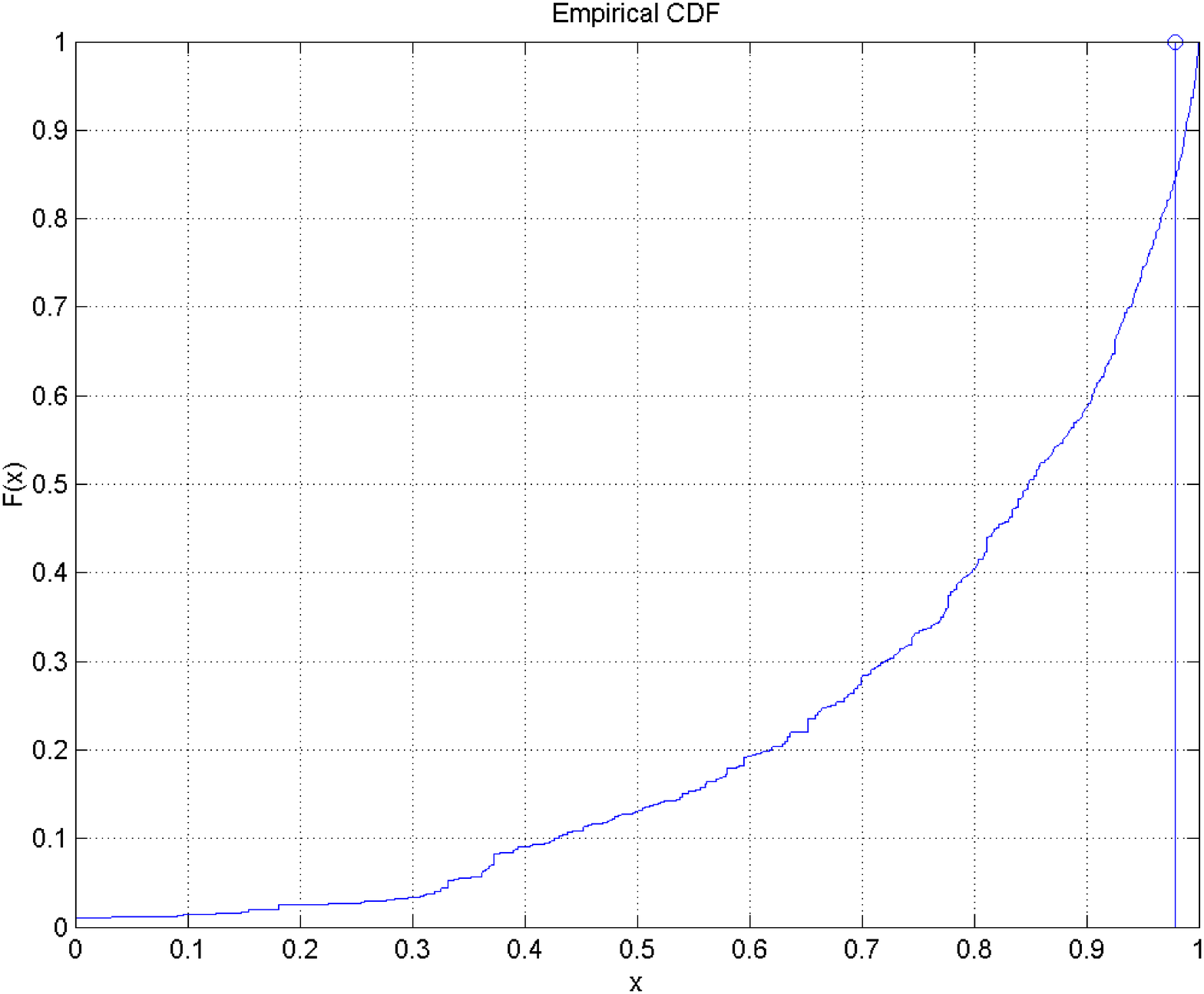}}
\caption{Cumulative Distribution Function for bootstrapped
$I(\textit{Pax2 motif:GTTCC} \rightarrow Y)$; $Y$ is the class label
(UB/non-UB). True $\hat{I} (\textit{Pax2} \rightarrow
\textit{Gata3}) = 0.9792$. }\label{fig:cdf_motif}
\end{figure}

\subsection{}
The other question again picks up on something quite traditionally
done in bioinformatics research - finding key TF regulators
underlying tissue-specific expression. Again, using the
\textit{Gata3} gene as an example, it has been observed that it is
expressed in the developing ureteric bud (UB) during kidney
development. To find UB specific regulators, we look for conserved
TF modules in the promoters of UB-specific genes. These
experimentally annotated UB-specific genes are obtained from the
Mouse Genome Informatics database at
\emph{http://www.informatics.jax.org/}. Several programs are used
for this kind of analysis, like Genomatix \cite{Genomatix} or Toucan
\cite{Toucan2005}. Using Toucan, we align the promoters of the
various UB specific genes, and obtained several related modules. The
top-ranking module in Toucan contains \textit{AHR-ARNT},
\textit{Hox13}, \textit{Pax2}, \textit{Tal1alpha-E47},
\textit{Oct1}. We can now check if the corresponding motifs can
discriminate UB-specific and non-specific genes, from DTI.

We can now check if the \textit{Pax2} binding motif (GTTCC
\cite{Dressler1992}) induces kidney specific expression by looking
for the strength of DTI between the GTTCC motif and the class label
($+1$) indicating UB expression. This once again adds to
computational evidence for the true role of \textit{Pax2} in
directing ureteric bud specific expression \cite{Dressler1992}.

\section{Discussion}
From the results above, we observe that the average
misclassification error is higher in the heart/brain promoter
datasets than for the enhancer data. We speculate this is due to a
number of reasons:
\begin{itemize}
\item
There is more sequence variability at the promoter since it has to
act in concert with LREs of different tissue types.
\item
Since the enhancer/LRE acts with the promoter to confer expression
in only one tissue type, these sequences are more specific and hence
their mining identifies motifs that are probably more indicative of
tissue-specific expression.
\end{itemize}

We however, reiterate that the enhancer dataset that we study always
make use of the \emph{hsp68-lacz} as the promoter driven by the
ultraconserved elements. Hence we do not have promoter specificity.
Though this is a disadvantage and might not reveal all key motifs,
it is the best that can be done in the absence of any other
comprehensive repository.

%


\section{Conclusions}
In this work, we have presented a framework for the identification
of hexamer motifs to discriminate between two kinds of sequences
(tissue-specific promoters vs non-specific or tissue-specific
regulatory elements vs non-specific).  For this feature selection
problem we proposed the utility of a new metric - the 'directed
information' (DTI). In conjunction with a support vector machine
classifier, this method was shown to outperform the state of the art
methods employing ordinary mutual information. We also find that
only a subset of the discriminating motifs correlate with known
transcription factor motifs and hence might be potentially related
to underlying epigenetic phenomena governing tissue-specific
expression. The superior performance of the directed-information
based variable selection suggests its utility to more general
sequential learning problems.

We also examine the applicability of DTI for questions focusing on
any motif of interest, obtained as output from other sources
(literature, expression data, module searches). Thus one can
prospectively resolve the role of a TF in a biological process.

\section{Acknowledgements}
The authors gratefully acknowledge the support of the NIH under
award 5R01-GM028896-21 (J.D.E). We would also like to thank Prof.
Sandeep Pradhan and Mr. Ramji Venkataramanan for useful discussions
on Directed information.



{\footnotesize
\begin{thebibliography}{}
\addtolength{\baselineskip}{-.1\baselineskip}

\bibitem{paper4}
Stuart RO, Bush KT, Nigam SK, "Changes in gene expression patterns
in the ureteric bud and metanephric mesenchyme in models of kidney
development", {\em Kidney International},64(6),1997-2008,December
2003.
\bibitem{DrosophilaFeatureDiff}
Chan BY, Kibler D.,Using hexamers to predict cis-regulatory motifs
in Drosophila.,BMC Bioinformatics. 2005 Oct 27;6:262.
\bibitem{icassp2006}
Rao,Hero AO,States DJ,Engel JD,"Inference of biologically relevant
Gene Influence Networks using the Directed Information
Criterion",{\em Proc. of the IEEE Conference on Acoustics, Speech
and Signal Processing}, 2006.
\bibitem{paper 24}
Casella G. and Berger RL. "Statistical Inference", {\em Duxbury
Press}, 1990.
\bibitem{PromFind}
Hutchinson GB.,"The prediction of vertebrate promoter regions using
differential hexamer frequency analysis".,{\em Comput Appl Biosci}.
1996 Oct;12(5):391-8.
\bibitem{paper 19}
NCBI Pubmed URL: {\em http://www.ncbi.nlm.nih.gov/entrez/query.fcgi}
\bibitem{Darbellay-Vajda1999}
G. A. Darbellay and I. Vajda, "Estimation of the information by an
adaptive partitioning of the observation space," {\em IEEE Trans. on
Information Theory}, vol. 45, pp. 1315--1321, May 1999.
\bibitem{HastieTibshirani2002}
Hastie T, Tibshirani R,The Elements of Statistical Learning ,
Springer 2002.
\bibitem{Fleuret2004}
F. Fleuret. "Fast binary feature selection with conditional mutual
information", {\em Journal of Machine Learning Research,},
1531–1555, 2004.
\bibitem{Guyon2004}
I. Guyon, A. Elisseeff, "An Introduction to Variable and Feature
Selection", {\em Journal of Machine Learning Research} 3 1157-1182,
\bibitem{Shimizu2006}
Kadota K, Ye J, Nakai Y, Terada T, Shimizu K.,"ROKU: a novel method
for identification of tissue-specific genes",.2003. {\em BMC
Bioinformatics}. 2006 Jun 12;7:294.
\bibitem{Stoeckert2005}
Schug , J., Schuller, W-P., Kappen, C., Salbaum, J.M., Bucan, M.,
Stoeckert, C.J. Jr.,"Promoter Features Related to Tissue Specificity
as Measured by Shannon Entropy"., {\em Genome Biology} 6(4): R33,
March 2005.
\bibitem{Geweke1982}
Geweke J.,"The Measurement of Linear Dependence and Feedback Between
Multiple Time Series," {\em Journal of the American Statistical
Association}, 1982, 77, 304-324.  (With comments by E. Parzen, D. A.
Pierce, W. Wei, and A. Zellner, and rejoinder)
\bibitem{CoD2003}
R. Hashimoto, E.R.Dougherty, M. Brun, Z. Zhou, M.L. Bittner, J.M.
Trent, "Efficient selection of feature-sets possessing hig
coefficients of determination based on incremental
determinations",{\em Signal Processing}. 83 (4) (2003) 695-712.
\bibitem{Ghosh2005}
Ghosh D, Chinnaiyan AM., "Classification and selection of biomarkers
in genomic data using LASSO".,{\em J Biomed Biotechnol}.,
2005(2):147-54.
\bibitem{PRECISE2005}
Trindade LM, van Berloo R, Fiers M,
Visser RG.,"PRECISE: software for prediction of cis-acting
regulatory elements".,{\em J Hered}. 2005 Sep-Oct;96(5):618-22.
\bibitem{NIPS2006}
Proc. NIPS 2006 Workshop on
Causality and Feature Selection, available at:
\emph{http://research.ihost.com/cws2006/}
\bibitem{Dressler1992}
Dressler, G.R. and Douglas, E.C. (1992)., "Pax-2 is a DNA-binding
protein expressed in embryonic kidney and Wilms tumor"., {\em Proc.
Natl. Acad. Sci. USA} 89: 1179-1183.
\bibitem{Massey1990}
J. Massey, "Causality, feedback and directed information," {\em
Proc. 1990 Symp. Information Theory and Its Applications
(ISITA-90)}, Waikiki, HI, Nov. 1990, pp. 303–305.
\bibitem{Pradhan2007}
Pradhan, S. S., "On the Role of Feedforward in Gaussian Sources:
Point-to-Point Source Coding and Multiple Description Source
Coding," Information Theory, {\em IEEE Transactions on Information
Theory} , vol.53, no.1pp.331-349, Jan. 2007.
\bibitem{Burge1997}
Burge C, Karlin S: "Prediction of complete gene structures in human
genomic DNA". {\em J Mol Biol} 1997, 268:78-94.
\bibitem{RPscore}
King DC, Taylor J, Elnitski L, Chiaromonte F, Miller W, Hardison
RC., "Evaluation of regulatory potential and conservation scores for
detecting cis-regulatory modules in aligned mammalian genome
sequences".,{\em Genome Res}. 2005 Aug;15(8):1051-60. Epub 2005 Jul
15.
\bibitem{EnhancerBrowser}
Pennacchio, L. A., Ahituv, N., Moses, A., Prabhakar, S., Nobrega,
M., Shoukry, M., Minovitsky, A., Dubchak, I., Holt, A., Lewis, K.,
Plazer-Frick, I., Akiyama, J., DeVal, S., Afzal, V., Black, B.,
Couronne, O., Eisen, M., Visel, A., and Rubin, E.M. 2006., "In vivo
enhancer analysis of human conserved non-coding sequences", {\em
Nature}, 444(7118):499-502.
\bibitem{Toucan2005}
Aerts S, Van Loo P, Thijs G, Mayer H, de Martin R, Moreau Y, De Moor
B.,TOUCAN 2: the all-inclusive open source workbench for regulatory
sequence analysis.,Nucleic Acids Res. 2005 Jul 1;33(Web Server
issue):W393-6.
\bibitem{Genomatix}
Werner T., "Regulatory networks: Linking microarray data to systems
biology".,{\em Mech Ageing Dev}. 2007 Jan;128(1):168-72.
\bibitem{Gubner2006}
Gubner J. A., Probability and Random Processes for Electrical and
Computer Engineers, Cambridge, 2006.
\bibitem{John1989b}
H. Joe., "Relative entropy measures of multivariate dependence".
{\em J. Am. Statist. Assoc}., 84:157–164, 1989.
\bibitem{Silverman1997}
J. Ramsay, B. W. Silverman, Functional Data Analysis (Springer
Series in Statistics), Springer 1997.
\bibitem{HastieTibshirani2002}
Hastie T, Tibshirani R, The Elements of Statistical Learning ,
Springer 2002.
\bibitem{EffronTibshirani1994}
Effron B, Tibshirani R.J, An Introduction to the Bootstrap
(Monographs on Statistics and Applied Probability), Chapman \&
Hall/CRC, 1994.
\bibitem{NemenmanBialek2002}
Nemenman, F Shafee, and W Bialek. "Entropy and inference,
revisited.", In TG Dietterich, S Becker, and Z Ghahramani, editors,
{\em Advances in Neural Information Processing Systems} 14,
Cambridge, MA, 2002. MIT Press.
\bibitem{WilettNowak2004}
Willett R, Nowak R, "Complexity-Regularized Multiresolution Density
Estimation", ISIT 2004.
\bibitem{Paninski2003}
Paninski, L. (2003)., "Estimation of entropy and mutual
information". {\em Neural Computation} 15: 1191-1254.
\bibitem{ErikLearnedMiller2003}
Erik Learned-Miller and John W. Fisher, III. "ICA using spacings
estimates of entropy". {\em Journal of Machine Learning Research
(JMLR)}, Volume 4, pp. 1271-1295, 2003.
\bibitem{SignalProcMag2006}
Hudson, J.E., "Signal Processing Using Mutual Information", {\em
Signal Processing Magazine},Volume: 23,no: 6 pp:50-54, Nov. 2006.
\bibitem{}
Visel A, Minovitsky S, Dubchak I, Pennacchio LA., "VISTA Enhancer
Browser--a database of tissue-specific human enhancers"., {\em
Nucleic Acids Res}. 2007 Jan;35(Database issue):D88-92.
\bibitem{Pennachio2007}
Pennacchio LA, Loots GG, Nobrega MA, Ovcharenko I., "Predicting
tissue-specific enhancers in the human genome"., {\em Genome Res}.
2007 Jan 8;
\bibitem{Pennachio2005}
Prabhakar S, Poulin F, Shoukry M, Afzal V, Rubin EM, Couronne O,
Pennacchio LA., "Close sequence comparisons are sufficient to
identify human cis-regulatory elements"., {\em Genome Res}. 2006
Jul;16(7):855-63.
\bibitem{Kleinjan2005}
Kleinjan DA, van Heyningen V., "Long-range control of gene
expression: emerging mechanisms and disruption in disease"., {\em Am
J Hum Genet}. 2005 Jan;76(1):8-32.
\bibitem{Khandekar2004}
Khandekar M, Suzuki N, Lewton J, Yamamoto M, Engel JD., "Multiple,
distant Gata2 enhancers specify temporally and tissue-specific
patterning in the developing urogenital system".,{\em Mol Cell
Biol}. 2004 Dec;24(23):10263-76.
\bibitem{CNSGata3}
Lakshmanan, G., K. H. Lieuw, K. C. Lim, Y. Gu, F. Grosveld, J. D.
Engel, and A. Karis. 1999. "Localization of distant urogenital
system-, central nervous system-, and endocardium-specific
transcriptional regulatory elements in the GATA-3 locus". {\em Mol.
Cell. Biol}. 19:1558-1568.
\bibitem{Lieb2006}
Lieb JD, Beck S, Bulyk ML, Farnham P, Hattori N, Henikoff S, Liu XS,
Okumura K, Shiota K, Ushijima T, Greally JM., "Applying whole-genome
studies of epigenetic regulation to study human disease".,{\em
Cytogenet Genome Res}. 2006;114(1):1-15.
\bibitem{Kreiman2004}
Kreiman G., "Identification of sparsely distributed clusters of
cis-regulatory elements in sets of co-expressed genes".,{\em Nucleic
Acids Res}. 2004 May 20;32(9):2889-900.
\bibitem{Fraenkel2006}
MacIsaac KD, Fraenkel E., "Practical strategies for discovering
regulatory DNA sequence motifs".,{\em PLoS Comput Biol}. 2006
Apr;2(4):e36.
\bibitem{MI_feature}
Peng H.,Long F.,Ding C.,"Feature Selection Based on Mutual
Information: Criteria of Max-Dependency, Max-Relevance, and
Min-Redundancy", {\em IEEE Transactions on Pattern Analysis and
Machine Intelligence}, vol. 27, No. 8, pp: 1226-1238, August 2005.
\end{thebibliography}
}
\end{document}